# Orbital Period Refinement of CoRoT Planets With TESS Observations

*Peter Klagyivik (1,2,3), Hans J. Deeg (2,3), Szilárd Csizmadia (1), Juan Cabrera (1) and Grzegorz Nowak (2,3)*

[1]*Deutsches Zentrum für Luft und Raumfahrt, Institut für Planetenforschung, Berlin, Germany,* [2]*Instituto de Astrofísica de Canarias (IAC), Tenerife, Spain,* [3]*Department of Astrofísica, Universidad de La Laguna (ULL), Tenerife, Spain*

CoRoT was the first space mission dedicated to exoplanet detection. Operational between 2007 and 2012, this mission discovered 37 transiting planets, including CoRoT-7b, the first terrestrial exoplanet with a measured size. The precision of the published transit ephemeris of most of these planets has been limited by the relative short durations of the CoRoT pointings, which implied a danger that the transits will become unobservable within a few years due to the uncertainty of their future transit epochs. Ground-based follow-up observations of the majority of the CoRoT planets have been published in recent years. Between Dec. 2018 and Jan. 2021, the TESS mission in its sectors 6 and 33 re-observed those CoRoT fields that pointed towards the Galactic anti-center. These data permitted the identification of transits from nine of the CoRoT planets, and the derivation of precise new transit epochs. The main motivation of this study has been to derive precise new ephemerides of the CoRoT planets, in order to keep these planets' transits observable for future generations of telescopes. For large expensive telescopes, this means that transit times should be known with a precision of better than 30 min. The TESS data were analyzed for the presence of transits and the epochs of these re-observed transits were measured. The original CoRoT epochs, epochs from ground-based follow-up observations and those from TESS were collected. From these data updated ephemerides are presented for nine transiting planets discovered by the CoRoT mission in its fields pointing towards the Galactic anti-center. In three cases (CoRoT-4b, 19b and 20b), transits that would have been lost for ground observations, due to the large uncertainty in the previous ephemeris, have been recovered. The updated ephemerides permit transit predictions with uncertainties of less than 30 min for observations at least until the year 2030. No significant transit timing variations were found in these systems.

## 1 INTRODUCTION

CoRoT was the first space mission dedicated to extrasolar planet detections (Baglin et al., 2006; Auvergne et al., 2009). The mission was active between 2008 and 2012 and covered 24 fields with pointings of varying lengths, between 24 and 153 days. All pointings were within one of the "CoRoT Eyes" - two circular zones with a radius of about 7°, centered either close to the Galactic center (at 18 h 50m, 0deg, in equatorial coordinates) or to the anti-center (at 6 h 50m, 0deg). Data from the mission have led to the discovery of 37 transiting planet systems (Deleuil et al., 2018)[1], with 21 systems in the center and 16 in the anti-center fields.

TABLE 1 | Transit times of CoRoT-1b.

| $T_c$ | $\sigma_{T_c}$ | Time-Sys | Source | Comment |
|---|---|---|---|---|
| 54,159.4532 | 0.0001 | HJD_UTC | Barge et al. (2008) | |
| 55,529.59727 | 0.00079 | HJD_UTC | ETD, 52 | Naves R |
| 55,582.41301 | 0.00072 | HJD_UTC | ETD, 53 | Muler G |
| 56,596.43685 | 0.00076 | HJD_UTC | ETD, 64 | Sokov E. N., Rusov S |
| 56,629.63404 | 0.00148 | HJD_UTC | ETD, 67 | Gonzalez J |
| 58,476.61293 | 0.00058 | BJD_TDB | TESS, this work | |
| 58,826.69423 | 0.00057 | HJD_UTC | ETD, 97 | Yves Jongen |
| 58,894.59651 | 0.00059 | HJD_UTC | ETD, 100 | Yves Jongen |
| 58,897.61539 | 0.00061 | HJD_UTC | ETD, 101 | Yves Jongen |
| 58,900.63236 | 0.0006 | HJD_UTC | ETD, 102 | Yves Jongen |
| 59,212.98925 | 0.00043 | BJD_TDB | TESS, this work | |
| 59,229.5872 | 0.00061 | HJD_UTC | ETD, 104 | Yves Jongen |
| 59,279.38296 | 0.00078 | HJD_UTC | ETD, 107 | Jens Jacobsen |

Given the length of the CoRoT pointings, the precision of the planet's ephemeris for the prediction of their future transit events is limited accordingly, especially for planets found in the shorter pointings and/or for faint targets with low signal-to-noise ratio. These planets might be practically "lost" for future ground observations, which happens when uncertainties of a transit prediction exceed three or 4 h, which makes a re-observation in a given observing night unfeasible. For the recovery of such "lost ephemerides", ground observations would need dedicated observing campaigns that cover all potential times of transit occurrence within the uncertainty, which implies a time-consuming dedicated observing effort, typically over multiple nights of potential transit occurrences. A space mission like TESS (Ricker et al., 2015) that provides uninterrupted coverage over a longer span - of 28 days in the case of TESS - is therefore much more efficient for ephemeris maintenance and recovery.

Long term monitoring of exoplanet transits may lead to the discovery of additional planets or stellar companions in the same system. Transit timing variations (hereafter TTV) are a reliable sign of gravitational perturbations caused by a third body in the planetary system. The first planet discovered with this method was Kepler-19c (Ballard et al., 2011). The amplitude of the difference between the predicted and observed transit times due to TTVs is usually in the range of a few minutes. This is in the range to be detectable with photometric observations from the ground or from a small-aperture space telescope like TESS, with increasing sensitivity towards the detection of TTVs from transit-timings that are acquired across longer spans in time. However, since TESS to date has observed the CoRoT-fields at only two epochs in winter 2018/19 and 2020/21, the CoRoT and TESS timings alone are not effective for TTV studies. Additional ground-based observations from the intermediate time-span between the discovery by CoRoT and the first TESS re-observation are therefore valuable.

After the end of the original photometric follow-up campaign accompanying the CoRoT mission (Deeg et al., 2009), which was primarily aimed towards the identification of false alarms among planet candidates, later efforts centered on the obtainment of updated ephemerides. To date, two dedicated works have been published: For one, the re-determination of transit ephemeris from ground-observation for six CoRoT planets by Raetz et al. (2019, hereafter R19), and for another, ground-based transit epochs for 20 CoRoT planets by Deeg et al. (2020), with five planets covered by both R19 and D20. Of note is also the collection of re-observations which are compiled in the Exoplanet Transit database (ETD Poddaný et al., 2010). These data are mainly acquired by amateurs; in many cases (see also comments in D20, **Table 1**) they are however of insufficient quality for a reliable re-determination of a new ephemeris.

The present work is a continuation of D20, where we extend the ground-based observations with space-based timings for those planets for which suitable light-curves have been acquired by the TESS mission.

CoRoT's Galactic center fields - in which a slight majority of its planets were found - were initially (before the TESS launch) included in the area covered by TESS Sectors 25 or 26. However, due to excessive contamination by stray Earth- and moonlight in TESS cameras 1 and 2, during subsequent operational preparations these sectors were shifted northwards, leaving these CoRoT fields without any coverage. The present scheduling for TESS does not foresee any coverage of this region either, due to which the center fields might be observed at earliest in Winter 2022/23.

TESS observed the CoRoT-fields towards the Galactic anti-center twice. First in its Sector six in winter 2018/9 and then in its Sector 33, 2 years later. The following work is therefore exclusively describing results on CoRoT planets in the anti-center fields. Another condition for inclusion in this work has been a successful identification of the planet's transits in TESS short-cadence data, which provide a sampling of 120 s. This led to the following sample included in this paper: CoRoT-1b, CoRoT-4b, CoRoT-5b, CoRoT-7b, CoRoT-12b, CoRoT-13b, CoRoT-18b, CoRoT-19b and CoRoT-20b.

In **Section 2** we present the observational data used for the analysis. Transit times are also listed together with the transit fitting method for the TESS observations. In **Section 3** we calculate the updated ephemeris and orbital period of the investigated planets and the predicted transit timing errors for the next decade. In **Section 4** we shortly discuss all planets one by one, while **Section 5** contains our conclusions.

---

[1]A full list of the CoRoT planets can be generated from Deleuil et al.'s electronic table A2, by selecting those entries with a 'planet' identifier.

## 2 OBSERVATIONAL DATA AND TRANSIT TIMINGS

In the following paragraphs we describe the different sources of the ephemeris or transit timings that are used in this work. In brief, the following sources of transit observations were used:

- the CoRoT ephemeris, mostly from the planets' discovery publications; • reliable ground-based follow-up observations; • TESS observations.

**Supplementary Table S1** in the supplementary material lists all transit times together with their sources and eventual comments. **Table 1** shows the content of **Supplementary Table S1** for the case of CoRoT-1b.

### 2.1 Ephemeris From CoRoT Planet Discoveries

For the original ephemeris (prior to this work), we used in most cases an ephemeris that is based exclusively on CoRoT data, given in a planet's discovery publication. There were two exceptions to this: For CoRoT-18b, the discovery ephemeris by Hébrard et al. (2011) took already into account a further ground-based timing, obtained 8 months after the CoRoT pointing. For CoRoT-7b, we don't consider the discovery ephemeris by Léger et al. (2009), but the much more precise one by Barros et al. (2014), which was derived after a second CoRoT pointing to that planet was performed in early 2012.

These discovery ephemerides have errors associated to their epoch and period values that were obtained by a variety of methods, depending on the choice of the author(s) that performed the work. Inconsistencies among these error-values were already noted during the first years of the ground-based follow-up, when it became evident that the errors of some planets were deviating by factors of "a few" from values that could be expected for a given system. Given the need for an estimator that reliably determines when - and if - follow-up transit observations should be scheduled, a method to estimate "standardized" ephemeris errors was then employed, which depends on basic parameters of the system, namely the depth and duration of its transits, the noise of the light curve, and the number of consecutive transits that are covered (Deeg, 2015; Deeg and Tingley, 2017).

Hence, in addition to the originally published ephemeris errors we indicate for all targets the homogeneous "standardized" ephemeris errors, obtained by this method[2]. The standardized epoch errors were calculated with Eq. 5 of Deeg and Tingley (2017), where we used the expected noise in CoRoT light curves from **Eq. (1)** in Aigrain et al. (2009), with the R magnitude of the target as input, and the transit depths and durations from the discovery publications. The standardized epoch error was then converted into a corresponding period error using Eq. 7 of Deeg (2015), which uses the number of consecutive transits in a given light curve as a further input. Note that we did not re-evaluate the published values of the epochs or periods themselves, but only their errors.

The CoRoT discovery ephemeris, with both their originally published and with their "standardized" errors, are indicated in **Table 2**. We note that all originally published CoRoT ephemeris were given in heliocentric UTC time, $HJD_{UTC}$, because the time-stamps of the original CoRoT light curves were given in that system[3]. For the analysis, we converted all previous ephemeris and transit times to $BJD_{TDB}$, hence the updated ephemeris in **Table 2** are also in $BJD_{TDB}$.

The epochs of the discovery ephemeris are also given in the list of transit timings (**Supplementary Table S1**), as the first entry for each planet.

### 2.2 Ground-Based Transit Timings

For the CoRoT planets with successful TESS transit identifications (see Sect. 2.3), **Supplementary Table S1** provides their published ground-based timings. These timings are mostly from the two prior works dedicated to CoRoT ground follow-up, namely R19 and D20. From ETD, whose timings are mostly sourced from amateur observers, we include those data that were found to be of sufficient reliability to be useful for the re-determination of the transit ephemeris reported in Sect. 3. These are generally timings with a Quality Index (DQ, as defined by Poddaný et al., 2010) of ≤3. One exception is CoRoT-1b, where there are over 90 timings presented in ETD. Here we used only the timings with DQ ≤ 2.

### 2.3 Transit Timings From TESS

As mentioned, at present TESS has covered only the CoRoT anti-center fields, with their corresponding planet sample. All of these planets were observed by TESS in its Sector six between 2018 Dec 11 and 2019 Jan 7, and in its Sector 33 between 2020 Dec 17 and 2021 Jan 13. However, not all of the CoRoT anti-center planets are included in this work. Only those planets published before the beginning of the TESS mission received dedicated target apertures in TESS observations, which implies the availability of light curves with a high temporal resolution (of 120 s; in Sector 33 also of 20 s). This led to the exclusion of two cases: CoRoT-15b (Bouchy et al., 2011), because it is not a planet but a brown dwarf and hence did not enter into the sample for TESS. CoRoT-37b (listed in Deleuil et al., 2018, but without a description; see also D20) was not included in the TESS sample because it is still lacking a formal publication of its discovery, due to which it is also excluded from current catalogs of exoplanets.

---

[2]An exception is CoRoT-7b, for which no standardized error has been calculated. The calculation of the standardized errors assumes a sequence of consecutive transits, which is not compatible with the ephemeris of CoRoT-7b by Barros et al. (2014), which is based on two CoRoT runs separated by 4 years; see also Sect. 4.4.

[3]The timestamps in the original CoRoT data releases (DATEJDHEL column, prior to Version 4) were in heliocentric UTC times. In the reprocessed 'N2 Legacy' data release, available since 2017 at the IAS Corot Public Archive, DATEJDHEL has been replaced by a Barycentric Dynamical Time scale BJDTDB, in the DATEBARTT column (Chaintreuil et al., 2016).

**TABLE 2** | Previous and updated ephemerides of the analyzed planets. The updated ephemerides are in $BJD_{TDB}$, while the previous ephemerides are in $HJD_{UTC}$. The previous ephemerides are given with errors from the original literature and with standardized errors (see text Sect. 2.1).

| $T_0$ | Period [days] | | Source |
|---|---|---|---|
| **CoRoT-1b** | | | |
| previous ephemeris | | | |
| 2454159.4532 ± 0.0001 | 1.5089557 ± 0.0000064 | original | Barge et al. (2008) |
| 2454159.4532 ± 0.0003 | 1.5089557 ± 0.0000027 | standardized | |
| updated ephemeris | | | |
| 2454159.4543 ± 0.0003 | 1.50896848 ± 0.00000011 | | this study |
| **CoRoT-4b** | | | |
| previous ephemeris | | | |
| 2454141.36416 ± 0.00089 | 9.20205 ± 0.00037 | original | Aigrain et al. (2008) |
| 2454141.36416 ± 0.00098 | 9.20205 ± 0.00015 | standardized | |
| updated ephemeris | | | |
| 2454141.36498 ± 0.00098 | 9.2016281 ± 0.0000040 | | this study |
| **CoRoT-5b** | | | |
| previous ephemeris | | | |
| 2454400.19885 ± 0.0002 | 4.0378962 ± 0.0000019 | original | Rauer et al. (2009) |
| 2454400.19885 ± 0.00098 | 4.0378962 ± 0.0000172 | standardized | |
| updated ephemeris | | | |
| 2454400.19974 ± 0.00071 | 4.0379148 ± 0.0000011 | | this study |
| **CoRoT-7b** | | | |
| previous ephemeris | | | |
| 2454398.07756 ± 0.0006 | 0.85359159 ± 0.00000057 | original | Barros et al. (2014) |
| updated ephemeris | | | |
| 2454398.0783 ± 0.0006 | 0.8535926 ± 0.0000009 | | this study |
| **CoRoT-12b** | | | |
| previous ephemeris | | | |
| 2454398.62707 ± 0.00036 | 2.828042 ± 0.000013 | original | Gillon et al. (2010) |
| 2454398.62707 ± 0.00138 | 2.828042 ± 0.000014 | standardized | |
| updated ephemeris | | | |
| 2454398.6279 ± 0.0012 | 2.8280517 ± 0.0000012 | | this study |
| **CoRoT-13b** | | | |
| previous ephemeris | | | |
| 2454790.8091 ± 0.0006 | 4.03519 ± 0.00003 | original | Cabrera et al. (2010) |
| 2454790.8091 ± 0.0024 | 4.03519 ± 0.00004 | standardized | |
| updated ephemeris | | | |
| 2454790.8105 ± 0.0023 | 4.0350906 ± 0.0000036 | | this study |
| **CoRoT-18b** | | | |
| previous ephemeris | | | |
| 2455321.72412 ± 0.00018 | 1.9000693 ± 0.0000028 | original | Hébrard et al. (2011) |
| 2455321.72412 ± 0.00046 | 1.9000693 ± 0.0000250 | standardized | |
| updated ephemeris | | | |
| 2455321.72504 ± 0.00039 | 1.90009057 ± 0.00000044 | | this study |
| **CoRoT-19b** | | | |
| previous ephemeris | | | |
| 2455257.44102 ± 0.0006 | 3.89713 ± 0.00002 | original | Guenther et al. (2012) |
| 2455257.44102 ± 0.0012 | 3.89713 ± 0.00018 | standardized | |
| updated ephemeris | | | |
| 2455257.4418 ± 0.0012 | 3.8971372 ± 0.0000021 | | this study |
| **CoRoT-20b** | | | |
| previous ephemeris | | | |
| 2455266.0001 ± 0.00135 | 9.24285 ± 0.0003 | original | Deleuil et al. (2012) |
| 2455266.0001 ± 0.0005 | 9.24285 ± 0.00034 | standardized | |
| updated ephemeris | | | |
| 2455266.0006 ± 0.0012 | 9.2431839 ± 0.0000072 | | this study |

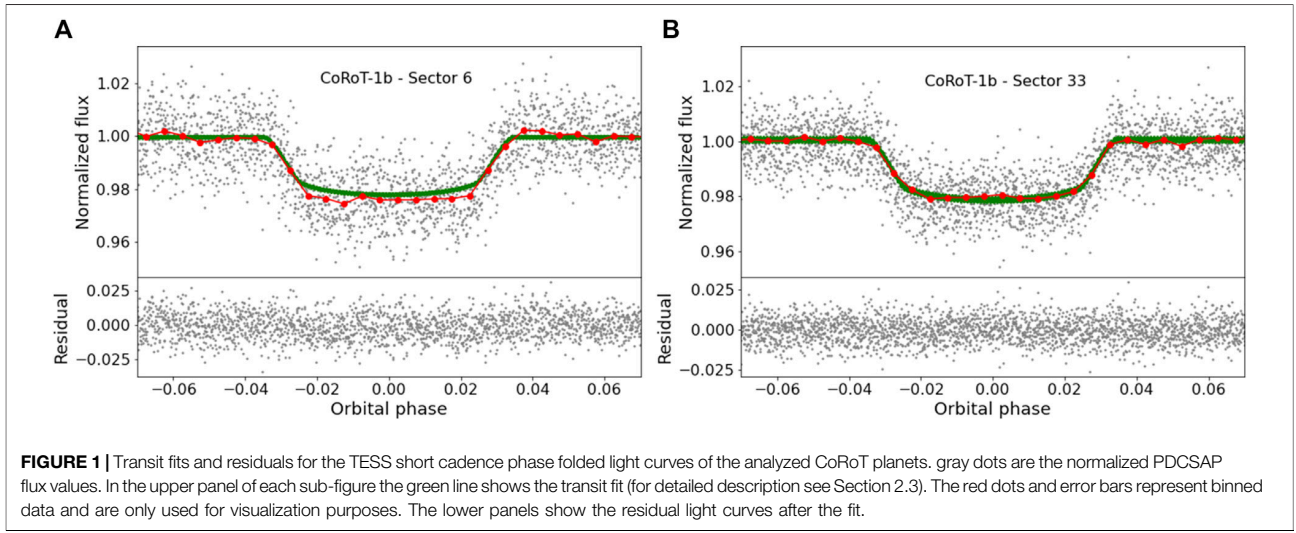

**FIGURE 1 |** Transit fits and residuals for the TESS short cadence phase folded light curves of the analyzed CoRoT planets. gray dots are the normalized PDCSAP flux values. In the upper panel of each sub-figure the green line shows the transit fit (for detailed description see Section 2.3). The red dots and error bars represent binned data and are only used for visualization purposes. The lower panels show the residual light curves after the fit.

We collected from the NASA's Mikulsky Archive for Space Telescopes (MAST)[4] all available TESS 120s light curves of the CoRoT anti-center planets, as well as the light curves from the 20s fast mode. Transits of CoRoT-21b (Pätzold et al., 2012), CoRoT-24b, CoRoT-24c (Alonso et al., 2014), CoRoT-31b (Bordé et al., 2020) and CoRoT-32b (Boufleur et al., 2018) were not detectable in the TESS short cadence light curves, — neither as individual transits, nor in light curves that had been folded by the known orbital periods. Non-detections of some CoRoT systems with faint host-stars, or with shallow transits, were expected, due to the much smaller telescope-aperture of TESS of 10 cm, versus the 27 cm effective aperture of CoRoT. For CoRoT-15b, which was not observed in short cadence mode, we also checked light curves obtained from full frame images (Huang et al., 2020), but we did not find any significant transit signal. We also note that CoRoT-31b (Bordé et al., 2020) was observed in short cadence mode only in Sector 6, but not in Sector 33.

From the downloaded light-curves, we used the "PDCSAP" fluxes, which are fluxes from on-board aperture photometry that had undergone a "Pre-search Data Conditioning" procedure (Jenkins et al., 2020), aimed at the suppression of signals that would be deterrent to the detection of transits.

Since the periods of the planets are known we folded the light curves using this period. Any possible period change since the publication date of the original value does not blur the folded light curve of the 28 days long observation and therefore it is negligible.

TESS obtained the light curves of CoRoT planets at lower signal-to-noise ratios than CoRoT. Therefore, instead of determining the times of all individual transits, we decided to determine only one mid-transit time of the studied planets by modeling all their transits during the TESS observations simultaneously. The assigned epoch is the one that corresponds to the central transit in the light curve of a TESS sector.

[4] https://archive.stsci.edu

We used the Transit and Light Curve Modeller (TLCM, Csizmadia 2020) to model the light curves and to obtain the mid-transit times. Other parameters, like planet-to-star radius ratio, impact parameter, eccentricity and argument of periastron, period, etc. had priors on the values given in previous publications (Aigrain et al., 2008; Barge et al., 2008; Cabrera et al., 2010; Gillon et al., 2010; Hébrard et al., 2011; Deleuil et al., 2012; Guenther et al., 2012; Raetz et al., 2019) and they could vary only between their reported ± 1$\sigma$ values. During the modeling, the quadratic limb darkening law was used. Gaussian priors for the limb darkening coefficients were taken from Claret (2017) and the width of the priors were defined by propagating the input stellar data into the values of the limb darkening coefficients.

TLCM utilized a spherical star and planet model (Mandel and Agol, 2002). After optimizing the fit with a Genetic Algorithm, the result was refined by three simultaneous Simulated Annealing chains. The error bars were estimated by MCMC-analysis. The several MCMC-chains were controlled by continuously monitoring the Gelman-Rubin statistics, the autocorrelation length of the chains and the estimated sample size (Croll 2006; Ford 2006; Foreman-Mackey et al., 2013). The chain length was automatically extended if convergence or a good mixing was not reached. The reported values in **Supplementary Table S1** are the median values of the posterior distributions and the reported 1$\sigma$ error bars are obtained by a common 16–84% rule.

In **Figure 1** we present the light curve of CoRoT-1b obtained by TESS and the corresponding model fit for both Sector 6 (A) and Sector 33 (B). The TESS light curves and model fits of all other targets are presented in **Supplementary Figures S1, S2** in the supplementary material.

## 3 UPDATED EPHEMERIS

As already mentioned in Sect 2.1, instead of the epochs of the individual transits from CoRoT observations, we used the epoch of the originally published ephemeris. Similarly, we derived a single transit epoch for each sector of TESS observations. On the other hand, the epochs from ground-based timings arise from the observations of single transits, and hence these epochs generated individual data-points.

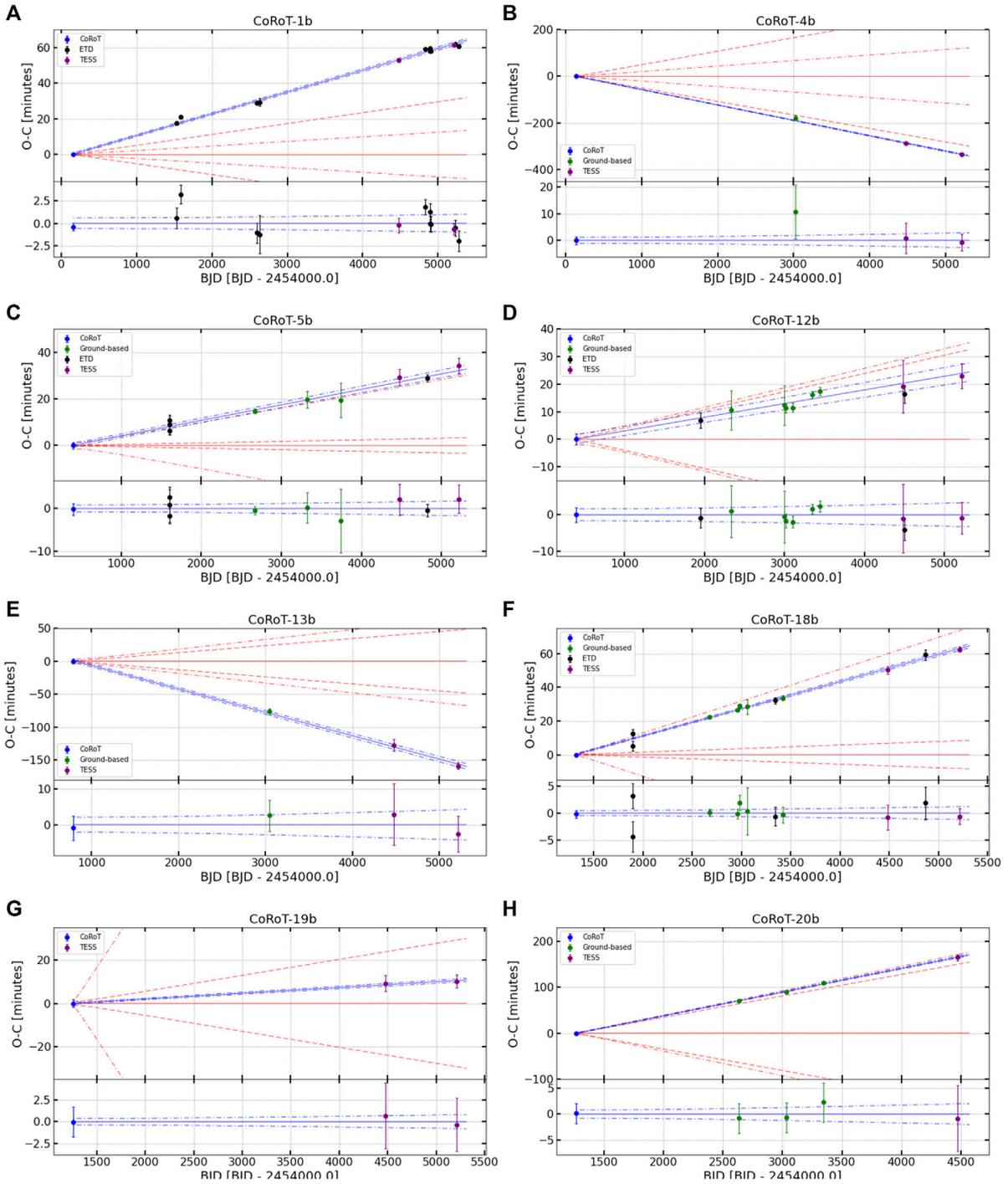

**FIGURE 2 |** O-C diagrams of the planets' transits, against the discovery ephemeris **(upper part of each panel)** and against the revised ephemeris **(lower part)**. In each panel, the blue dot on the left is the CoRoT measurement, green dots represent any ground-based observation, while the purple dots correspond to TESS observations in Sectors 6 and 33. The red lines refer to timing predictions based on the original ephemeris, where the dashed and the dash-dotted lines correspond to the original and the standard timing errors, respectively. The blue solid line represents the updated ephemeris, with its errors shown as dash-dotted lines.

TABLE 3 | Current and future uncertainties of transit epochs, from original, standardized and updated ephemerides, for the indicated dates.

| Planet | orig. Ephem | St. ephem | Updated ephem | Updated ephem | Updated ephem |
|---|---|---|---|---|---|
| | January 01, 2022 [min] | January 01, 2022 [min] | January 01, 2022 [min] | January 01, 2030 [min] | January 01, 2200 [min] |
| CoRoT-1b | 33.1 | 14.0 | 0.7 | 1.0 | 7.4 |
| CoRoT-4b | 314.9 | 128.6 | 3.7 | 5.4 | 44 |
| CoRoT-5b | 3.5 | 31.8 | 2.3 | 3.4 | 28 |
| CoRoT-7b | 5.3 | - | 8.2 | 12.8 | 110 |
| CoRoT-12b | 34.3 | 36.9 | 3.7 | 5.4 | 44 |
| CoRoT-13b | 51.3 | 71.4 | 7.0 | 10.4 | 90 |
| CoRoT-18b | 9.0 | 80.6 | 1.5 | 2.4 | 23 |
| CoRoT-19b | 32.0 | 292.8 | 3.8 | 6.0 | 55 |
| CoRoT-20b | 201.7 | 230.0 | 5.2 | 8.3 | 78 |

In the upper panels of **Figure 2** we plot O-C times of the epochs from **Supplementary Table S1** against the planets' original ephemeris (see **Table 2**). Furthermore, the lower panels of **Figure 2** show O-C times against newly-derived ephemeris of increased precision. These revised ephemerides were derived from linear fits to the same epochs, minimizing the O-C times. These fits are weighted by the errors of the individual timings. We note that the errors of both the CoRoT and TESS epochs are based on the number of transits on which they are based. A simple weighted fit is therefore suitable and will not under-represent the importance of the CoRoT or TESS transits. The updated ephemerides are also included in **Table 2**.

One of the major aims of this study has been to keep the CoRoT planets's transits observable in the future, through the provision of reliable ephemerides. **Table 3** shows the uncertainties in the prediction of transit times for the beginning of the year 2022, from the original ephemeris based on the published and the standardized ephemeris errors. Using the revised ephemeris from this work, timing uncertainties for the current (2022) and several future epochs are indicated as well.

## 4 DISCUSSION

For any detailed future studies of the CoRoT planets' transits, e.g. for transmission spectroscopy to characterize their atmospheres, precise ephemerides are crucial for their scheduling.

In this paper, we have revised nine planetary systems discovered by CoRoT that were observed by NASA's TESS mission.

The improvements in timing precision (comparing here the precisions for 2022 in **Table 3**) are between one and two orders of magnitude. An exception is CoRoT-7b, where the precision did not increase, because the original ephemeris was already rather precise, due to having been based on two separate CoRoT pointings, and TESS detected the transit only marginally.

In the following, we provide further details to each of the planets in the sample.

TABLE 4 | RV observations of CoRoT-1b.

| BJD_TDB | RV [km/s] | Source |
|---|---|---|
| 2454184.306625 | 23.2879 ± 0.0385 | Barge et al. (2008) |
| 2454185.310225 | 23.1284 ± 0.0600 | Barge et al. (2008) |
| 2454192.304915 | 23.5625 ± 0.0271 | Barge et al. (2008) |
| 2454197.321395 | 23.2862 ± 0.0309 | Barge et al. (2008) |
| 2454376.665429 | 23.5784 ± 0.0226 | Barge et al. (2008) |
| 2454378.663959 | 23.3730 ± 0.0324 | Barge et al. (2008) |
| 2454379.665409 | 23.6200 ± 0.0224 | Barge et al. (2008) |
| 2454380.670939 | 23.6907 ± 0.0229 | Barge et al. (2008) |
| 2454381.632369 | 23.3795 ± 0.0227 | Barge et al. (2008) |
| 2459268.440505 | 24.0238 ± 0.0204 | this study |
| 2459268.454971 | 23.9989 ± 0.0228 | this study |
| 2459268.466246 | 23.7104 ± 0.0253 | this study |
| 2459269.360419 | 23.5367 ± 0.0234 | this study |
| 2459269.374886 | 23.4342 ± 0.0228 | this study |

### 4.1 CoRoT-1b

The timing error against the original (Barge et al., 2008) ephemeris was about 29 min on January 01, 2020. Neither R19 nor D20 performed any ground-based observations, however, there are over 90 transit timings reported in ETD. Using all transits with DQ ≤ 3 results in a large scatter with outliers up to 20 min, therefore we decided to use only the best quality observations with DQ ≤ 2. The TESS transits appeared to be ∼1 h late, which is a 2$\sigma$ and 4$\sigma$ difference compared to the prediction using the original and the standardized error, respectively. The ETD timings follow the same trend and gives an updated ephemeris consistent with our results. The difference in both $T_0$ and orbital period is within 4$\sigma$ compared to our results. Our updated ephemeris gives a timing error of 1.0 min for January 01, 2030.

A radial velocity drift of about 1 m/s per day was reported for CoRoT-1b from RVs obtained at two epochs in March-April 2007 and in October 2007, at the SOPHIE spectrograph of the 193 cm telescope at the Observatoire Haute Provence, France (Barge et al., 2008; Bonomo et al., 2017; Csizmadia, 2020). Assuming that this drift is real, its influence onto an otherwise linear ephemeris can be expressed by (Deeg et al., 2008, Eq. 6)

$$(O - C)_E = E^2 a \frac{P^2}{2c}, \quad (1)$$

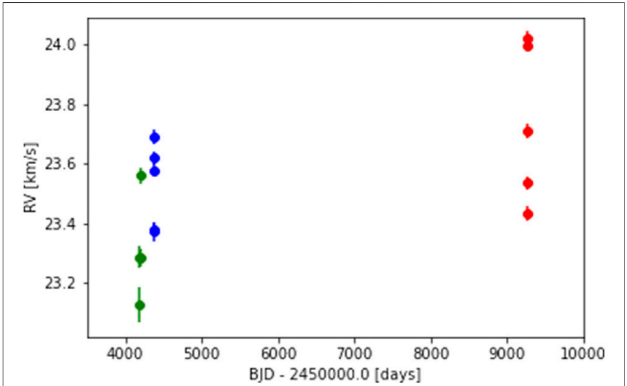

**FIGURE 3** | Absolute radial velocities of CoRoT-1b. Data from the three observing runs are shown by different colors.

where $E$ is the epoch number (number of orbits in time-span under consideration), $a$ the system's acceleration in the line of sight, $P$ the orbital period, and $c$ the speed of light.

With the system acceleration given by $a = 1/86400$ m/$s^2$ = $1.2 \, 10^{-5}$ m/$s^2$, the orbital period of 1.51 days and extrapolating over E = 3,000 orbits (accounting from mid 2007 to early 2020), we obtain a delay of $(O-C)_E \approx 3,100$ s, which agrees rather well with the above mentioned lateness of the TESS transits by 1 h. Motivated by this finding, which might imply the presence of a longer-periodic third body, we obtained further RV data on the nights of 22 and Feb. 23, 2021 with the FIES echelle spectrograph on the Nordic Optical Telescope under the Spanish CAT observing programme 59–210. The data were acquired with the high-res 1.3 arcsec fibre offering a spectral resolution of R = 67,000. The RV standard HD 3765 was observed in both nights with the same setup, and the data were reduced with the FIES pipeline based on CERES (Brahm et al., 2017). The results are included in **Table 4**. These new data have a relatively large scatter but clearly indicate that the previous rate of RV drift is not maintained. In order to derive a new value, we subtract the effect of Corot-1b from all measured RVs, using the updated ephemeris given in **Table 2** and the RV amplitude of K = 188 m/s for a circular orbit, given by (Barge et al., 2008). For the measurements in each of the three observing epochs, March/april 2007, October 2007 and February 2021, we then derive systemic RVs which are 23.36, 23.55 and 23.70 km/s, respectively, all with a formal error of ±0.02 km/s. From a linear fit through these systemic RVs, over the ≈14 years covered by them, we obtain now a much smaller average RV drift of only 0.05 m/s per day, much below the previous estimates. This implies either that the previously reported large RV drift was dominated by zero-point errors among the two initial observing campaigns, or that the RV drift from 2007 is real, but has a periodicity that is significantly smaller than the 14 years that passed until the next observations. Further RV monitoring of the system would be therefore be needed to determine the amplitude and periodicity of RV variations caused by any potential longer-periodic planet in this system.

### 4.2 CoRoT-4b

For CoRoT-4b the timing error on January 01, 2020 was already 272.6 min. Fortunately D20 managed to observe a transit with the IAC80 telescope, but the transit occurred 3 hours earlier than predicted. This means a 2.5$\sigma$ deviation from the standardized predicted transit error. However, the transit times obtained by TESS are in very good agreement with the ground-based data, which suggests that the discrepancy is caused by the inaccuracy of the original period value. The updated ephemeris results in a timing error of 5.4 min for January 01, 2030.

### 4.3 CoRoT-5b

Since the timing error of CoRoT-5b was 27.4 min on January 01, 2020, we did not plan any ground based observations. However, R19 published three transits. All these transits are 11$\sigma$ later than predicted, however, the calculated standardized error is 10× larger than the original value published in the discovery paper (Rauer et al., 2009) and hence the errors seem to be underestimated. The TESS 2 min cadence observations follow the same trend as the ground based transits, including the ETD timings. The updated ephemeris gives a timing error of 3.4 min for January 01, 2030.

### 4.4 CoRoT-7b

CoRoT-7b was the first rocky planet discovered by the transit method (Léger et al., 2009). Its transit depth is only 0.03%, which is too small for any ground-based follow-up. CoRoT-7b is however also the only CoRoT planet that was observed in two CoRoT runs, namely in run LRa01 from October 24, 2007 to March 3, 2008, resulting in its initial discovery reported by Léger et al. (2009), and in a later run (LRa06, from January 10, 2012 to March 29, 2012) that was specifically dedicated to this planet (Barros et al., 2014), and which led to the original ephemeris reported in **Table 2**.

We searched blind for the transit signal of CoRoT-7b in TESS data and we failed to get a significant detection. The transit detection statistic DST (Cabrera et al., 2012) did produce some excess of signal above the noise at the expected period of the planet in Sector 6, but not in Sector 33, and in any case it was not significant. However, combining the data from both sectors still produced some excess of signal, still not significant, consistent with a coherent transiting planet (a constant period across the two sectors, separated by several hundredths of days). This fact encouraged us to treat the excess of signal as a genuine transiting planet and we found that it has properties (orbital period, epoch, transit depth, transit duration) consistent with the expected signal [as characterized in Barros et al. (2014)]. We used pycheops (Maxted et al., in prep.) to model the signal and found the parameters below, which are roughly consistent with the expectations [as per Barros et al. (2014)]. We cannot improve the planetary parameters with TESS (the uncertainties are too large), but we can confirm that the we can detect the signal of CoRoT-7b with the expected properties.

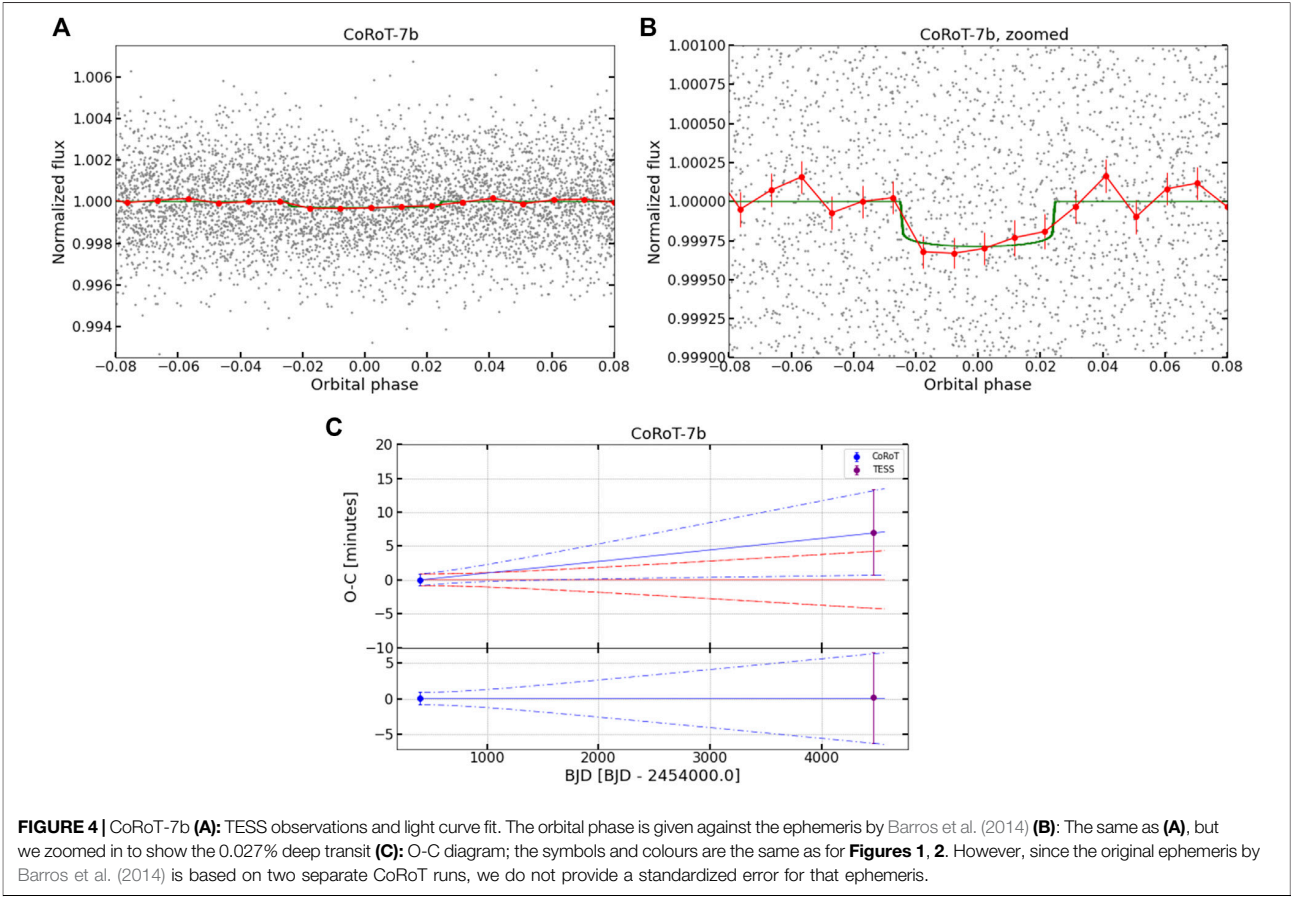

**FIGURE 4** | CoRoT-7b **(A):** TESS observations and light curve fit. The orbital phase is given against the ephemeris by Barros et al. (2014) **(B):** The same as **(A)**, but we zoomed in to show the 0.027% deep transit **(C):** O-C diagram; the symbols and colours are the same as for **Figures 1**, **2**. However, since the original ephemeris by Barros et al. (2014) is based on two separate CoRoT runs, we do not provide a standardized error for that ephemeris.

In the two top panels of **Figure 4** we present the combined Sector six and Sector 33 TESS short cadence light curve of CoRoT-7b (gray dots) together with the binned data (red points) and the transit fit (green line). The left panel shows the full phase folded light curve around the transit, while in the right panel we zoomed in to show the 0.027% deep transit.

Unfortunately, Barros et al. (2014) do not provide a separate epoch for the transits from the second CoRoT run. In the O-C diagram (bottom panel of **Figure 4**) we can therefore show only one CoRoT epoch, corresponding to the first run in late 2007, whereas the second CoRoT epoch would have been near a value of 2000 on the time axis. Nevertheless, our TESS timing agrees rather well with the Barros et al. ephemeris, deviating by only a few minutes, which also implies that there are no notable variations in orbital period between the time spans of 2007–2012 and 2012–2018.

The updated timing error is 12.8 min for 2030.

### 4.5 CoRoT-12b

For CoRoT-12b the ground-based observations are within both the one sigma predicted ephemeris error of the discovery paper (Gillon et al., 2010) and the standardized error. The TESS observations are in good agreement with the previous data points. As the target is close to the fainter limit of TESS, the ±10 min error of the transit timing is not sufficient for any small amplitude TTV analysis, which would otherwise be possible based on the ground-based observations with small error bars. The updated ephemeris error for 2030 is 5.4 min, which is a factor of 9 better than the originally predicted error calculated from the discovery paper.

### 4.6 CoRoT-13b

D20 detected a 76 min ($2\sigma$) early transit versus the ephemeris of the discovery paper (Cabrera et al., 2010), indicating potential TTVs. However, TESS observations are in perfect agreement with the ground-based data, which rather simply implies period uncertainty. The updated timing error is 10.4 min for January 01, 2030.

### 4.7 CoRoT-18b

Based on only the CoRoT observations the standardized timing error of CoRoT-18b on January 01, 2020 is already 66.8 min, however, the published error in the discovery paper (Hébrard et al., 2011) is six times smaller. Four ground-based follow-up transits were published by R19. All these transits are within the $1\sigma$ standardized error. Including our IAC80 observation and the TESS short cadence light curves, the updated ephemeris gives a timing error of 2.4 min for January 01, 2030.

### 4.8 CoRoT-19b

Since the standardized timing error of CoRoT-19b on January 01, 2020 is already 243.3 min, practically it has been lost for targeted observations. We detected a possible egress with IAC80, but the detection is not clear, therefore this observation is not included in the updated ephemeris and we used only the TESS short cadence light curve. The updated ephemeris results in a timing error of 6.0 min for January 01, 2030, which means that the TESS data saved this planet for future observations.

### 4.9 CoRoT-20b

This target was already lost for follow-up observations, since the timing error on January 01, 2020 was already 191 min. Fortunately R19 successfully observed two transits and D20 observed an additional one from the ground. Including the transits observed by TESS, the updated ephemeris gives a timing error of 8.3 min for January 01, 2030.

Rey et al. (2018) discovered the $17 \pm 1 M_{Jup}$ brown dwarf CoRoT-20c in the system. Its orbital period is $4.59 \pm 0.05$ years and it has an orbital eccentricity of $0.60 \pm 0.03$. They estimated the expected transit-timing variations of CoRoT-20b due to CoRoT-20c and found that the amplitude should be less than 5 min at a 68% credible interval. The confirmation of the TTV seems to be challenging both from the ground and using TESS data. However, there is a hint of transit duration variation, the TESS transits seem to be shorter than the CoRoT transits. Although this is only suspected, the photometric precision of the TESS data is not sufficient for accurate modeling. Further high precision photometric follow-up observations are needed to confirm this hypothesis. However, due to the brown dwarf companion and the fact that both the planet and the companion have highly eccentric orbit, it would not be surprising to detect such transit duration variation.

## 5 CONCLUSION

Due to the short observing run of CoRoT and the consequent period inaccuracy, for most planets the predicted transit timing error already reached several hours, in some cases even days. This error is dominated by the error of the period.

Thanks to the extensive ground-based follow-up activity and the TESS observations, the timing errors could be reduced to the order of a few minutes for the next 10 years, which is accurate enough to effectively use the observing time of the next generation telescopes (ELT, TMT, JWST, etc.) for atmospheric or any other type of characterization.

As can be noted in **Table 3**, the discovery ephemeris with their errors had already severe timing uncertainties for current observations of several planets (4, 19, 20b). A timing uncertainty exceeding 3 h would made their scheduling on ground based facilities very risky, with a high chance to lose a night of observation from waiting for a transit that does not occur, or occurs only partially, within that night.

The revised ephemerides, on the other hand, have sufficient precision to perform transit observations without any relevant loss of observing time, at least for the coming 1–2 decades. But these ephemerides–assuming that there are no relevant changes in the orbital period–provide also a precision that makes them useful over very long terms. Only in the year 2200, the planet with the least precise ephemeris, CoRoT-7b, would have accumulated an uncertainty that makes its scheduling in a given night somewhat problematic, but still feasible. Before such long terms come into play, we may however expect that most of these planets (and also the CoRoT center planets, which are not yet covered by TESS), might get further observing coverage by TESS[5] and then by future missions. The only predictable one is currently the PLATO mission (Rauer et al., 2014), scheduled to start scientific operations in 2027 and which will likely cover these fields during its second "step and stare" operational phase, giving coverages of a few weeks' duration.

In any case, the timings reported in this work, and especially those from TESS, will be relevant for the determination of any period variations among the observed planets, since the original CoRoT observations were too short to provide relevant constraints for these.

We would like to draw particular attention to CoRoT-1b. Although the radial velocity drift reported earlier is not present any more, the new RV observations suggest that there may be a third body in the system with an orbital period of less than 10 years. Therefore long-term radial velocity monitoring of CoRoT-1b is highly recommended.

## DATA AVAILABILITY STATEMENT

The original contributions presented in the study are included in the article/**Supplementary Material**, further inquiries can be directed to the corresponding author.

## AUTHOR CONTRIBUTIONS

PK and HD contributed to the conception and design of the study and wrote the manuscript. SC did all the light curve modeling and wrote the corresponding paragraphs. JC made the analysis of CoRoT-7b. GN made the additional spectroscopic observations and data analysis of CoRoT-1b. All authors contributed to manuscript revision, read, and approved the submitted version.

## FUNDING

The research by PK was carried out in part under the projects AP3400 Planetenforschung and Big-Data-Platform by the German Aerospace Center (DLR). PK and HD acknowledge

---

[5]At present, TESS observations are only scheduled to TESS Sector 55 in January 2022, without any coverage of CoRoT observing fields. Some coverage at later dates may however be expected.

support by grant ESP 2017-87676-C5-4-R of the Spanish Secretary of State for R&D&i (MINECO)

## ACKNOWLEDGMENTS

HD acknowledges also support from the Spanish Research Agency of the Ministry of Science and Innovation (AEI-MICINN) under grant "Contribution of the IAC to the PLATO Space Mission" with reference PID 2019-107061GB-C66, DOI: 10.13039/501100011033. We thank the referees for their useful comments.

## SUPPLEMENTARY MATERIAL

The Supplementary Material for this article can be found online at: https://www.frontiersin.org/articles/10.3389/fspas.2021.792823/full#supplementary-material

## REFERENCES

Aigrain, S., Collier Cameron, A., Ollivier, M., Pont, F., Jorda, L., Almenara, J. M., et al. (2008). Transiting Exoplanets from the CoRoT Space mission. IV. CoRoT-Exo-4b: a Transiting Planet in a 9.2 Day Synchronous Orbit. *Astron. Astrophysics* 488, L43–L46. doi:10.1051/0004-6361:200810246

Aigrain, S., Pont, F., Fressin, F., Alapini, A., Alonso, R., Auvergne, M., et al. (2009). Noise Properties of the CoRoT Data. A Planet-Finding Perspective. *Astron. Astrophysics* 506, 425–429. doi:10.1051/0004-6361/200911885

Alonso, R., Moutou, C., Endl, M., Almenara, J. M., Guenther, E. W., Deleuil, M., et al. (2014). Transiting Exoplanets from the CoRoT Space mission. XXVI. CoRoT-24: a Transiting Multiplanet System. *Astron. Astrophysics* 567, A112. doi:10.1051/0004-6361/201118662

Auvergne, M., Bodin, P., Boisnard, L., Buey, J. T., Chaintreuil, S., Epstein, G., et al. (2009). The CoRoT Satellite in Flight: Description and Performance. *Astron. Astrophysics* 506, 411–424. doi:10.1051/0004-6361/200810860

Baglin, A., Auvergne, M., Barge, P., Deleuil, M., Catala, C., Michel, E., et al. (2006). "Scientific Objectives for a Minisat: CoRoT,"in *The CoRoT Mission Pre-launch Status - Stellar Seismology and Planet Finding*. Editors M. Fridlund, A. Baglin, J. Lochard, and L. Conroy (of ESA Special Publication), 1306, 33.

Ballard, S., Fabrycky, D., Fressin, F., Charbonneau, D., Desert, J.-M., Torres, G., et al. (2011). The Kepler-19 System: A Transiting 2.2 R$_\oplus$ Planet and a Second Planet Detected via Transit Timing Variations. *Astrophysical J.* 743, 200. doi:10.1088/0004-637X/743/2/200

Barge, P., Baglin, A., Auvergne, M., Rauer, H., Léger, A., Schneider, J., et al. (2008). Transiting Exoplanets from the CoRoT Space mission. I. CoRoT-Exo-1b: a Low-Density Short-Period Planet Around a G0V star. *Astron. Astrophysics* 482, L17–L20. doi:10.1051/0004-6361:200809353

Barros, S. C. C., Almenara, J. M., Deleuil, M., Diaz, R. F., Csizmadia, S., Cabrera, J., et al. (2014). Revisiting the Transits of CoRoT-7b at a Lower Activity Level. *Astron. Astrophysics* 569, A74. doi:10.1051/0004-6361/201423939

Bonomo, A. S., Desidera, S., Benatti, S., Borsa, F., Crespi, S., Damasso, M., et al. (2017). The GAPS Programme with HARPS-N at TNG . XIV. Investigating Giant Planet Migration History via Improved Eccentricity and Mass Determination for 231 Transiting Planets. *Astron. Astrophysics* 602, A107. doi:10.1051/0004-6361/201629882

Bordé, P., Díaz, R. F., Creevey, O., Damiani, C., Deeg, H., Klagyivik, P., et al. (2020). Transiting Exoplanets from the CoRoT Space mission. XXIX. The Hot Jupiters CoRoT-30 B and CoRoT-31 B. *Astron. Astrophysics* 635, A122. doi:10.1051/0004-6361/201732393

Bouchy, F., Deleuil, M., Guillot, T., Aigrain, S., Carone, L., Cochran, W. D., et al. (2011). Transiting Exoplanets from the CoRoT Space mission. XV. CoRoT-15b: a Brown-dwarf Transiting Companion. *Astron. Astrophysics* 525, A68. doi:10.1051/0004-6361/201015276

Boufleur, R. C., Emilio, M., Janot-Pacheco, E., Andrade, L., Ferraz-Mello, S., do Nascimento, J., et al. (2018). A Modified CoRoT Detrend Algorithm and the Discovery of a New Planetary Companion. *Monthly Notices R. Astronomical Soc.* 473, 710–720. doi:10.1093/mnras/stx2187

Brahm, R., Jordán, A., and Espinoza, N. (2017). *CERES: A Set of Automated Routines for Echelle Spectra*, 129. Publications of the Astronomical Society of the Pacific. doi:10.1088/1538-3873/aa5455

Cabrera, J., Bruntt, H., Ollivier, M., Díaz, R. F., Csizmadia, S., Aigrain, S., et al. (2010). Transiting Exoplanets from the CoRoT Space mission . XIII. CoRoT-13b: a Dense Hot Jupiter in Transit Around a star with Solar Metallicity and Super-solar Lithium Content. *Astron. Astrophysics* 522, A110. doi:10.1051/0004-6361/201015154

Cabrera, J., Csizmadia, S., Erikson, A., Rauer, H., and Kirste, S. (2012). A Study of the Performance of the Transit Detection Tool DST in Space-Based Surveys. Application of the CoRoT Pipeline to Kepler Data. *Astron. Astrophysics* 548, A44. doi:10.1051/0004-6361/201219337

Chaintreuil, S., Deru, A., Baudin, F., Ferrigno, A., Grolleau, E., Romagnan, R., et al. (2016). *II.4 the "Ready to Use" CoRoT Data*, 61. doi:10.1051/978-2-7598-1876-1.c024

Claret, A. (2017). Limb and Gravity-Darkening Coefficients for the TESS Satellite at Several Metallicities, Surface Gravities, and Microturbulent Velocities. *Astron. Astrophysics* 600, A30. doi:10.1051/0004-6361/201629705

Croll, B. (2006). *Markov Chain Monte Carlo Methods Applied to Photometric Spot Modeling*, 118. Publications of the Astronomical Society of the Pacific, 1351–1359. doi:10.1086/507773

Csizmadia, S. (2020). The Transit and Light Curve Modeller. *Monthly Notices R. Astronomical Soc.* 496, 4442–4467. doi:10.1093/mnras/staa349

Deeg, H. J., Gillon, M., Shporer, A., Rouan, D., Stecklum, B., Aigrain, S., et al. (2009). Ground-based Photometry of Space-Based Transit Detections: Photometric Follow-Up of the CoRoT mission. *Astron. Astrophysics* 506, 343–352. doi:10.1051/0004-6361/200912011

Deeg, H. J., Klagyivik, P., Armstrong, J. D., Nespral, D., Tal-Or, L., Alonso, R., et al. (2020). Maintaining the Ephemeris of 20 CoRoT Planets: Transit Minimum Times and Potential Transit Timing Variations. *J. Am. Assoc. Variable Star Observers* 48, 201.

Deeg, H. J., Ocaña, B., Kozhevnikov, V. P., Charbonneau, D., O'Donovan, F. T., and Doyle, L. R. (2008). Extrasolar Planet Detection by Binary Stellar Eclipse Timing: Evidence for a Third Body Around CM Draconis. *Astron. Astrophysics* 480, 563–571. doi:10.1051/0004-6361:20079000

Deeg, H. J. (2015). Period, Epoch, and Prediction Errors of Ephemerides from Continuous Sets of Timing Measurements. *Astron. Astrophysics* 578, A17. doi:10.1051/0004-6361/201425368

Deeg, H. J., and Tingley, B. (2017). TEE, an Estimator for the Precision of Eclipse and Transit Minimum Times. *Astron. Astrophysics* 599, A93. doi:10.1051/0004-6361/201629350

Deleuil, M., Aigrain, S., Moutou, C., Cabrera, J., Bouchy, F., Deeg, H. J., et al. (2018). Planets, Candidates, and Binaries from the CoRoT/Exoplanet Programme. The CoRoT Transit Catalogue. *Astron. Astrophysics* 619, A97. doi:10.1051/0004-6361/201731068

Deleuil, M., Bonomo, A. S., Ferraz-Mello, S., Erikson, A., Bouchy, F., Havel, M., et al. (2012). Transiting Exoplanets from the CoRoT Space mission. XX. CoRoT-20b: A Very High Density, High Eccentricity Transiting Giant Planet. *Astron. Astrophysics* 538, A145. doi:10.1051/0004-6361/201117681

Ford, E. B. (2006). Improving the Efficiency of Markov Chain Monte Carlo for Analyzing the Orbits of Extrasolar Planets. *Astrophysical J.* 642, 505–522. doi:10.1086/500802

Foreman-Mackey, D., Hogg, D. W., Lang, D., and Goodman, J. (2013). *Emcee: The MCMC Hammer*, Publications of the Astronomical Society of the Pacific, 306. doi:10.1086/670067

Gillon, M., Hatzes, A., Csizmadia, S., Fridlund, M., Deleuil, M., Aigrain, S., et al. (2010). Transiting Exoplanets from the CoRoT Space mission. XII. CoRoT-12b: a Short-Period Low-Density Planet Transiting a Solar Analog star. *Astron. Astrophysics* 520, A97. doi:10.1051/0004-6361/201014981

Guenther, E. W., Díaz, R. F., Gazzano, J. C., Mazeh, T., Rouan, D., Gibson, N., et al. (2012). Transiting Exoplanets from the CoRoT Space mission. XXI. CoRoT-19b: a Low Density Planet Orbiting an Old Inactive F9V-star. *Astron. Astrophysics* 537, A136. doi:10.1051/0004-6361/201117706

Hébrard, G., Evans, T. M., Alonso, R., Fridlund, M., Ofir, A., Aigrain, S., et al. (2011). Transiting Exoplanets from the CoRoT Space mission. XVIII. CoRoT-18b: a Massive Hot Jupiter on a Prograde, Nearly Aligned Orbit. *Astron. Astrophysics* 533, A130. doi:10.1051/0004-6361/201117192

Huang, C. X., Vanderburg, A., Pál, A., Sha, L., Yu, L., Fong, W., et al. (2020). Photometry of 10 Million Stars from the First Two Years of TESS Full Frame Images: Part II. *Res. Notes Am. Astronomical Soc.* 4, 206. doi:10.3847/2515-5172/abca2d

Jenkins, J. M., Tenenbaum, P., Seader, S., Burke, C. J., McCauliff, S. D., Smith, J. C., et al. (2020). *Kepler Data Processing Handbook: Transiting Planet Search Kepler Science Document KSCI-19081-003*.

Léger, A., Rouan, D., Schneider, J., Barge, P., Fridlund, M., Samuel, B., et al. (2009). Transiting Exoplanets from the CoRoT Space mission. VIII. CoRoT-7b: the First Super-earth with Measured Radius. *Astron. Astrophysics* 506, 287–302. doi:10.1051/0004-6361/200911933

Mandel, K., and Agol, E. (2002). Analytic Light Curves for Planetary Transit Searches. *Astrophysical J. Lett.* 580, L171–L175. doi:10.1086/345520

Pätzold, M., Endl, M., Csizmadia, S., Gandolfi, D., Jorda, L., Grziwa, S., et al. (2012). Transiting Exoplanets from the CoRoT Space mission. XXIII. CoRoT-21b: a Doomed Large Jupiter Around a Faint Subgiant star. *Astron. Astrophysics* 545, A6. doi:10.1051/0004-6361/201118425

Poddaný, S., Brát, L., and Pejcha, O. (2010). Exoplanet Transit Database. Reduction and Processing of the Photometric Data of Exoplanet Transits. *New Astron.* 15, 297–301. doi:10.1016/j.newast.2009.09.001

Raetz, S., Heras, A. M., Fernández, M., Casanova, V., and Marka, C. (2019). Transit Analysis of the Corot-5, Corot-8, Corot-12, Corot-18, Corot-20, and Corot-27 Systems with Combined Ground- and Space-Based Photometry. *Monthly Notices R. Astronomical Soc.* 483, 824–839.

Rauer, H., Catala, C., Aerts, C., Appourchaux, T., Benz, W., Brandeker, A., et al. (2014). The PLATO 2.0 mission. *Exp. Astron.* 38, 249–330. doi:10.1007/s10686-014-9383-4

Rauer, H., Queloz, D., Csizmadia, S., Deleuil, M., Alonso, R., Aigrain, S., et al. (2009). Transiting Exoplanets from the CoRoT Space mission. VII. The "hot-Jupiter"-type Planet CoRoT-5b. *Astron. Astrophysics* 506, 281–286. doi:10.1051/0004-6361/200911902

Rey, J., Bouchy, F., Stalport, M., Deleuil, M., Hébrard, G., Almenara, J. M., et al. (2018). Brown dwarf Companion with a Period of 4.6 Yr Interacting with the Hot Jupiter CoRoT-20 B. *Astron. Astrophysics* 619, A115. doi:10.1051/0004-6361/201833180

Ricker, G. R., Winn, J. N., Vanderspek, R., Latham, D. W., Bakos, G. Á., Bean, J. L., et al. (2015). Transiting Exoplanet Survey Satellite (TESS). *J. Astronomical Telescopes, Instr. Syst.* 1, 014003. doi:10.1117/1.JATIS.1.1.014003

# *Orbital period refinement of the CoRoT planets with TESS observations – Supplementary Material*

## 1 SUPPLEMENTARY TABLES AND FIGURES

### 1.1 Tables

In the main part of our study we already presented the transit times of CoRoT-1b. Here, in Table S1 we list the transit times of all other targets.

### 1.2 Figures

As well as for the transit times, here we present the transit fits and residuals for the TESS short cadence phase folded light curves of the analysed CoRoT planets. Again, CoRoT-1b can be found in the main article as an example, and CoRoT-7b is also discussed separately in the main article in Section 5 of the main article. In Figures S1 and S2. Grey dots are the normalized PDCSAP flux values. In the upper panel of each sub-figure the green line shows the transit fit (for detailed description see Section 2.3 of the main article). The red dots and their error bars represent binned data and are only used for visualization purposes. The lower panels show the residual light curves after the fit.

## REFERENCES

Aigrain, S., Collier Cameron, A., Ollivier, M., Pont, F., Jorda, L., Almenara, J. M., et al. (2008). Transiting exoplanets from the CoRoT space mission. IV. CoRoT-Exo-4b: a transiting planet in a 9.2 day synchronous orbit. *Astronomy & Astrophysics* 488, L43–L46. doi:10.1051/0004-6361:200810246

Barros, S. C. C., Almenara, J. M., Deleuil, M., Diaz, R. F., Csizmadia, S., Cabrera, J., et al. (2014). Revisiting the transits of CoRoT-7b at a lower activity level. *Astronomy & Astrophysics* 569, A74. doi:10.1051/0004-6361/201423939

Cabrera, J., Bruntt, H., Ollivier, M., Díaz, R. F., Csizmadia, S., Aigrain, S., et al. (2010). Transiting exoplanets from the CoRoT space mission . XIII. CoRoT-13b: a dense hot Jupiter in transit around a star with solar metallicity and super-solar lithium content. *Astronomy & Astrophysics* 522, A110. doi:10.1051/0004-6361/201015154

Deeg, H. J., Klagyivik, P., Armstrong, J. D., Nespral, D., Tal-Or, L., Alonso, R., et al. (2020). Maintaining the Ephemeris of 20 CoRoT Planets: Transit Minimum Times and Potential Transit Timing Variations. *J. American Assoc. Variable Star Observers* 48, 201

Deleuil, M., Bonomo, A. S., Ferraz-Mello, S., Erikson, A., Bouchy, F., Havel, M., et al. (2012). Transiting exoplanets from the CoRoT space mission. XX. CoRoT-20b: A very high density, high eccentricity transiting giant planet. *Astronomy & Astrophysics* 538, A145. doi:10.1051/0004-6361/201117681

Gillon, M., Hatzes, A., Csizmadia, S., Fridlund, M., Deleuil, M., Aigrain, S., et al. (2010). Transiting exoplanets from the CoRoT space mission. XII. CoRoT-12b: a short-period low-density planet transiting a solar analog star. *Astronomy & Astrophysics* 520, A97. doi:10.1051/0004-6361/201014981

Guenther, E. W., Díaz, R. F., Gazzano, J. C., Mazeh, T., Rouan, D., Gibson, N., et al. (2012). Transiting exoplanets from the CoRoT space mission. XXI. CoRoT-19b: a low density planet orbiting an old inactive F9V-star. *Astronomy & Astrophysics* 537, A136. doi:10.1051/0004-6361/201117706

Hébrard, G., Evans, T. M., Alonso, R., Fridlund, M., Ofir, A., Aigrain, S., et al. (2011). Transiting exoplanets from the CoRoT space mission. XVIII. CoRoT-18b: a massive hot Jupiter on a prograde, nearly aligned orbit. *Astronomy & Astrophysics* 533, A130. doi:10.1051/0004-6361/201117192

**Table S1.** Transit times with their errors, the corresponding time system and the sources are presented. D20 refers to Deeg et al. (2020), R19 means Raetz et al. (2019) and ETD denotes the Exoplanet Transit Database (Poddaný et al., 2010).

| $T_c$ | $\sigma_{T_c}$ | Time-Sys. | Source | Comment |
|---|---|---|---|---|
| colspan=5 CoRoT-4b | | | | |
| 54141.36416 | 0.00089 | HJD_UTC | Aigrain et al. (2008) | |
| 57021.481 | 0.007 | BJD_UTC | D20 | |
| 58475.3323 | 0.0040 | BJD_TDB | TESS, this study | |
| 59211.4615 | 0.0022 | BJD_TDB | TESS, this study | |
| colspan=5 CoRoT-5b | | | | |
| 54400.19885 | 0.0002 | HJD_UTC | Rauer et al. (2009) | |
| 55599.45843 | 0.00123 | HJD_UTC | ETD, 2 | Schteinman G. |
| 55599.46027 | 0.00288 | HJD_UTC | ETD, 3 | Lomoz F. |
| 55603.49943 | 0.00114 | HJD_UTC | ETD, 4 | Schteinman G. M. |
| 56665.4696 | 0.0007 | BJD_TDB | R19 | |
| 57323.6502 | 0.0025 | BJD_TDB | R19 | |
| 57743.5912 | 0.0052 | BJD_TDB | R19 | |
| 58470.4193 | 0.0025 | BJD_TDB | TESS, this study | |
| 58825.75323 | 0.00104 | HJD_UTC | ETD, 10 | Yves Jongen |
| 59213.3957 | 0.0023 | BJD_TDB | TESS, this study | |
| colspan=5 CoRoT-7b | | | | |
| 2454398.07756 | 0.0006 | HJD_UTC | Barros et al. (2014) | |
| 2458468.8615 | 0.0044 | BJD_TDB | TESS, this study | |
| colspan=5 CoRoT-12b | | | | |
| 54398.62707 | 0.00036 | HJD_UTC | Gillon et al. (2010) | |
| 55948.39884 | 0.0019 | HJD_UTC | ETD, 2 | Garcia R. |
| 56997.606 | 0.005 | BJD_UTC | D20 | |
| 57014.5746 | 0.0013 | BJD_TDB | R19 | |
| 57099.4151 | 0.0011 | BJD_UTC | D20 | |
| 57342.6308 | 0.0009 | BJD_TDB | R19 | |
| 57444.4412 | 0.0011 | BJD_TDB | R19 | |
| 58479.5058 | 0.0066 | BJD_TDB | TESS, this study | |
| 58496.47124 | 0.0021 | HJD_UTC | ETD, 7 | Sjoerd Dufoer |
| 59211.9713 | 0.0031 | BJD_TDB | TESS, this study | |
| colspan=5 CoRoT-13b | | | | |
| 54790.8091 | 0.0006 | HJD_UTC | Cabrera et al. (2010) | |
| 57046.428 | 0.003 | BJD_UTC | D20 | |
| 58478.8854 | 0.0060 | BJD_TDB | TESS, this study | |
| 58531.42557 | 0.00596 | HJD_UTC | ETD, 2 | F. Lomoz |
| 59213.2680 | 0.0035 | BJD_TDB | TESS, this study | |
| colspan=5 CoRoT-18b | | | | |
| 55321.72412 | 0.00018 | HJD_UTC | Hébrard et al. (2011) | |
| 55893.65372 | 0.00161 | HJD_UTC | ETD, 2 | Lopesino J. |
| 55895.54858 | 0.00198 | HJD_UTC | ETD, 3 | Lopesino J. |
| 56678.3898 | 0.00046 | BJD_TDB | R19 | |
| 56959.603001 | 0.00059 | BJD_TDB | R19 | |
| 56978.605317 | 0.001 | BJD_TDB | R19 | |
| 57056.507 | 0.003 | BJD_UTC | D20 | |
| 57339.62002 | 0.00121 | HJD_UTC | ETD, 7 | Dauchet Coulon Charrier |
| 57419.424813 | 0.001 | BJD_TDB | R19 | |
| 58479.6750 | 0.0016 | BJD_TDB | TESS, this study | |
| 58865.39449 | 0.00208 | HJD_UTC | ETD, 11 | Manfred Raetz |
| 59213.1101 | 0.0010 | BJD_TDB | TESS, this study | |
| colspan=5 CoRoT-19b | | | | |
| 55257.44102 | 0.0006 | HJD_UTC | Guenther et al. (2012) | |
| 58476.4776 | 0.0026 | BJD_TDB | TESS, this study | |
| 59213.0358 | 0.0021 | BJD_TDB | TESS, this study | |
| colspan=5 CoRoT-20b | | | | |
| 55266.0001 | 0.00135 | HJD_UTC | Deleuil et al. (2012) | |
| 56633.9903 | 0.002 | BJD_UTC | D20 | |
| 57031.448 | 0.002 | BJD_TDB | R19 | |
| 57345.7182 | 0.0026 | BJD_TDB | R19 | |
| 58473.3842 | 0.0044 | BJD_TDB | TESS, this study | |

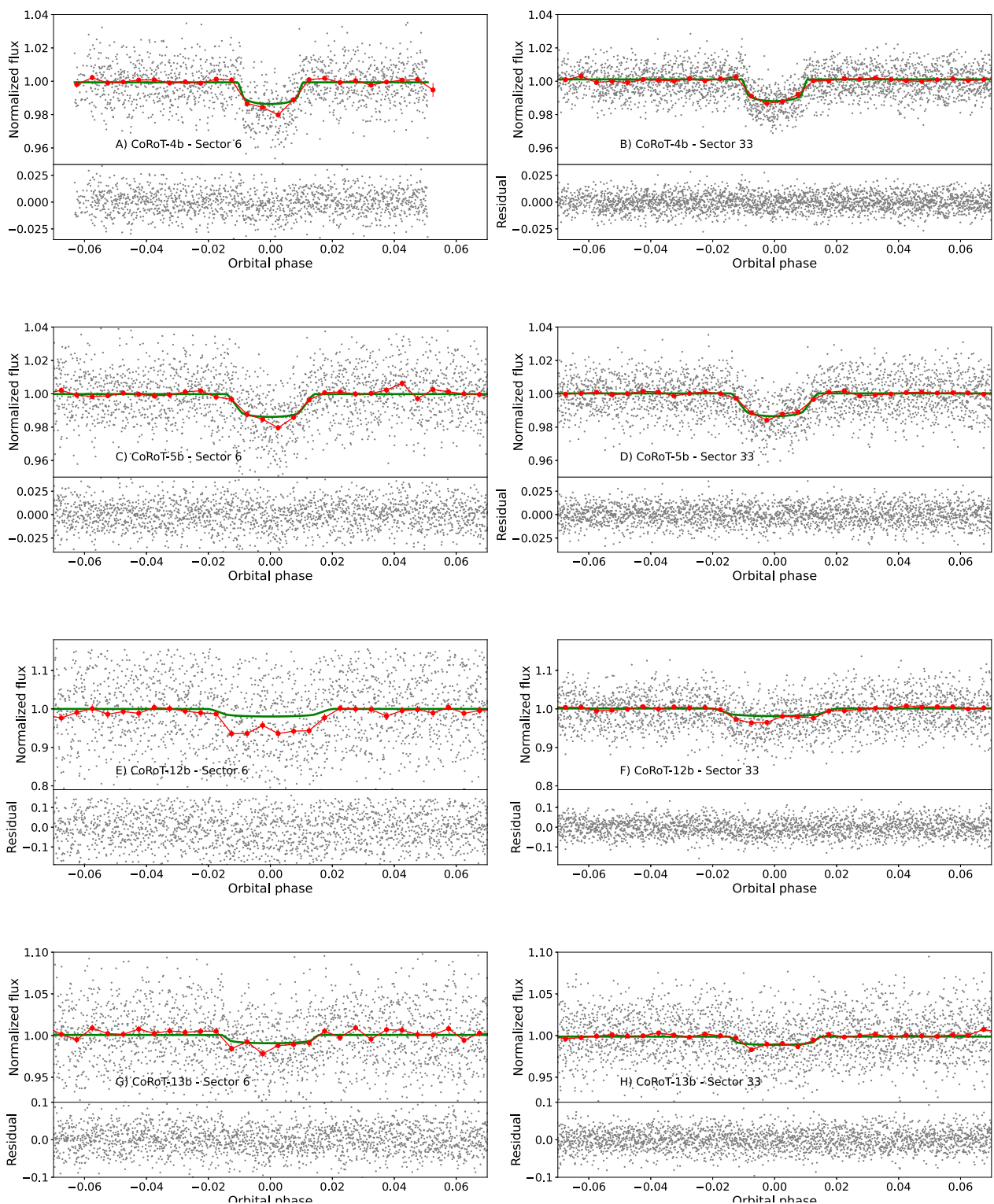

**Figure S1.** Transit fits and residuals for the TESS short cadence phase folded light curves of CoRoT-4b, CoRoT-5b, CoRoT-12b and CoRoT-13b.

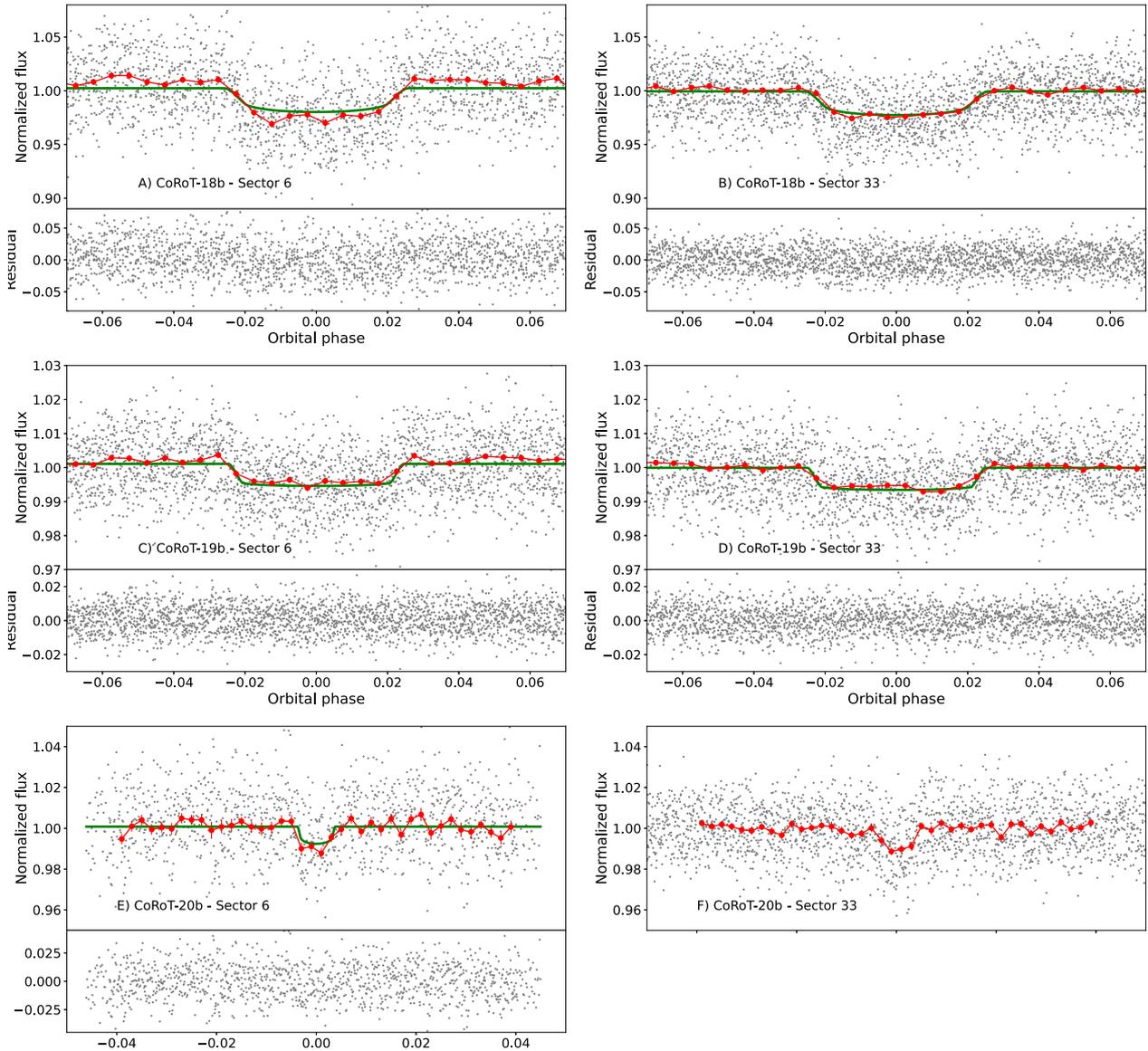

**Figure S2.** Transit fits and residuals for the TESS short cadence phase folded light curves of CoRoT-18b, CoRoT-19b, CoRoT-20b. **Although there is a clear transit of CoRoT-20b in Sector 33, the model fit did not converge. Therefore we did not use these data for our analysis.**

Poddaný, S., Brát, L., and Pejcha, O. (2010). Exoplanet Transit Database. Reduction and processing of the photometric data of exoplanet transits. *New Astronomy* 15, 297–301. doi:10.1016/j.newast.2009.09.001

Raetz, S., Heras, A. M., Fernández, M., Casanova, V., and Marka, C. (2019). "transit analysis of the corot-5, corot-8, corot-12, corot-18, corot-20, and corot-27 systems with combined ground- and space-based photometry". *Monthly Notices of the Royal Astronomical Society* 483, 824–839. doi:https://ui.adsabs.harvard.edu/abs/2019MNRAS.483..824R

Rauer, H., Queloz, D., Csizmadia, S., Deleuil, M., Alonso, R., Aigrain, S., et al. (2009). Transiting exoplanets from the CoRoT space mission. VII. The "hot-Jupiter"-type planet CoRoT-5b. *Astronomy &*

*Astrophysics* 506, 281–286. doi:10.1051/0004-6361/200911902